\documentclass[aps,pre,twocolumn,amsmath,amssymb,showpacs,superscriptaddress]{revtex4-1}% APS journal style

\usepackage{amsbsy}
\usepackage{amscd}
\usepackage{amsmath}
\usepackage{amsopn}
\usepackage{amstext}
\usepackage{amsxtra}
\usepackage{color}
\usepackage{enumerate}
\usepackage{epsfig}
\usepackage{hyperref}

\hypersetup{colorlinks,citecolor=blue,linkcolor=blue,filecolor=blue,urlcolor=blue}

\def\al{\alpha}
\def\bt{\beta}
\def\gm{\gamma}
\def\dl{\delta}

\def\nn{\nonumber}

\def\sg{\sigma}

\def\wh{\widehat}
\def\zt{\zeta}

\begin{document}
\title{Cascades on clique-based graphs}
\author{Adam Hackett}
\affiliation{Hamilton Institute, National University of Ireland, Maynooth, Co. Kildare, Ireland.}
\author{James P. Gleeson}
\affiliation{MACSI, Department of Mathematics \& Statistics, University of Limerick, Co. Limerick, Ireland.}
\date{\today}

% 89.75.Hc = Networks and genealogical trees
% 64.60.aq = Networks
% 64.60.ah = Percolation
% 87.23.Ge = Dynamics of social systems

\begin{abstract}
We present an analytical approach to determining the expected cascade size in a broad range of dynamical
models on the class of highly-clustered random graphs introduced by Gleeson [J. P. Gleeson, Phys. Rev. E \textbf{80},
036107 (2009)]. A condition for the existence of global cascades is also derived. Applications of this approach
include analyses of percolation, and Watts's model. We show how our techniques can be used to study the effects
of in-group bias in cascades on social networks.
\end{abstract}

\maketitle

%***********************************
\section{Introduction} \label{sec:1}
%***********************************

Emergent behaviors in a complex system depend crucially on the pattern of interactions between its components
\cite{barrat:08,easley:10,newman:10}. For example, we observe a cascade when local interactions in the vicinity
of an initially isolated effect allow that effect to propagate globally \cite{watts:02,gleeson:08}. The network
substrate of a system represents this pattern in its most abstract and analytically tractable form. This
information can be used to construct network models, which provide theoretical insights into the causes of such
behaviors. A fundamental problem for the construction of these models is the determination of precisely which
structural features are requisite to explain the phenomenon in question and which others are superfluous.

In the configuration model \cite{bender:78,bollobas:80} an ensemble of random graphs is prescribed by a degree
distribution $p_{k}$. In each realization drawn from this ensemble, a randomly selected vertex will have $k$
incident edges with probability $p_{k}$. This distribution represents the first order of complexity for most
network models. From this, a more realistic model can be constructed by including degree-degree correlations
\cite{newman:01,vazquez:03,goltsev:08,gleeson:10} and/or various forms of clustering
\cite{trapman:07,miller:09,newman:09,gleeson:09a}, both of which are explicitly absent from the configuration
model. Recently, the study of multiplex networks has introduced a further degree of complexity to this general
approach \cite{gao:11,brummitt:12,huang:13}. These networks consist of connected layers of networks, where each
layer involves interactions of a fundamentally unique kind.

In this paper we focus on random graphs with clustering; specifically, those defined by Gleeson in
\cite{gleeson:09a}. Real networks typically contain a large number of short cycles in which a small set of
vertices maintain a closed loop of connections. One way to measure the propensity for a vertex to form these
types of bonds is through the local clustering coefficient, which is defined as the fraction of pairs of
neighbors of a vertex that are also neighbors of each other \cite{watts:98}. The degree-dependent clustering
coefficient or clustering spectrum $c_k$ is found by averaging the local clustering coefficient over the class
of vertices of degree $k$ \cite{vazquez:02,serrano:06}. A global measure of clustering $C_2$ can be defined by
averaging the local coefficients of all $N$ vertices in the graph. Gleeson \cite{gleeson:09a} has shown how the
configuration model can be modified to generate ensembles of highly clustered graphs (see also
\cite{trapman:07,miller:09,newman:09}). This is achieved by embedding cliques of connected vertices into an
otherwise treelike structure. Each ensemble is prescribed by the joint distribution $\gm(k,c)$: the probability
that in any realization a randomly selected vertex has degree $k$ and is in a clique of $c$
vertices (a $c$-clique).

Our aim is to provide a generalized analytical approach to determining the expected cascade size on these
$\gm(k,c)$ or clique-based graphs. This goes far beyond the bond percolation process studied in 
\cite{gleeson:09a} to include a broad class of cascade processes including Watts's
threshold model \cite{watts:02}, \emph{k}-core decomposition \cite{goltsev:06,dorogovtsev:06}, and 
both site and bond percolation \cite{broadbent:57,stauffer:92}. Also of relevance is our earlier work 
\cite{hackett:11} on cascades on edge-triangle graphs \cite{newman:09,miller:09}. Edge-triangle graphs 
are created by embedding $3$-cliques, and only $3$-cliques, into an otherwise treelike structure. In each 
such graph a randomly chosen vertex is incident to $s$ single edges and $2t$ triangle edges with 
probability $p(s,t)$. In contrast, a $\gm(k,c)$ graph can contain cliques of many different sizes and 
may therefore have local clustering levels that are much higher than those in edge-triangle graphs. 
Furthermore, $\gm(k,c)$ can be parametrized to match the empirical clustering spectrum $c_k$ 
and degree distribution $p_k$ of a real-world network \cite{gleeson:09a}. This additional complexity 
means that a very different analytical approach from that of \cite{hackett:11} is required here. Our 
approach thus provides another significant extension of the methods used by Gleeson and Cahalane 
\cite{gleeson:07} and Gleeson \cite{gleeson:08}, who provided analytical results for cascades on 
configuration-model graphs by introducing a tree-based framework of level-by-level vertex activations. 
This method was inspired by methods originally developed to study the zero-temperature random-field Ising 
model on a Bethe lattice \cite{sethna:93,dhar:97,shukla:03}.

The class of cascade dynamics examinable through the tree-based framework consists of those processes that
satisfy the following list of properties: (i) each vertex is assigned a binary value specifying its current
state, active (damaged or infected) or inactive (undamaged or susceptible); (ii) the probability of a vertex 
becoming active (in a synchronous update of all vertices) depends only on its degree $k$ and the number $m$ 
of its neighbors that are already active and is termed the neighborhood influence response function 
$F_{m}^{k}$ \cite{watts:07,lopez:08}; (iii) for any fixed degree $k$, $F_{m}^{k}$ is a nondecreasing function 
of $m$; and (iv) once active, a vertex cannot become deactivated \footnote{An extension of this approach to 
nonmonotone binary-state dynamics has been provided in \cite{gleeson:11}.}. Each of the processes referred to 
in the preceding paragraph satisfies these constraints and is defined by choosing an appropriate $F_{m}^{k}$, 
as detailed in \cite{gleeson:08}. The goal of our analytical approach is the prediction of the expected size 
of the cascade when a time-dependent process of the type described here has run to completion. Our analytical 
results are defined as the fixed point of an iterative process, i.e., the solution of a self-consistent system 
of equations, but the level-by-level activation approach used in our analysis should not be misunderstood as a 
time-dependent process in its own right; rather it is a convenient representation of the iteration scheme for 
solving for the steady-state solution.

The remainder of this paper is structured as follows. In Sec.~\ref{sec:2} we describe in broad outline our
generalized approach to cascade dynamics on clique-based graphs. As well as an analytical expression for the
expected cascade size, we provide a first-order condition for the existence of cascades whose size scales with
the number of vertices $N$ as $N\rightarrow\infty$. Section~\ref{sec:3} deals in greater detail with the
particulars of clique member activations. We show how to calculate in closed form the number of active vertices
in a clique of any size $c\leq k+1$. The analysis of both sections is described in terms of an arbitrary
response function. The particular forms that this response takes for various processes are discussed in
Sec.~\ref{sec:4}, where we demonstrate the correspondence between our analytical results and numerical
simulations of bond percolation and Watts's model. In Sec.~\ref{sec:5} we present a possible extension of 
Watts's model in which different weights are assigned to active clique neighbors and active nonclique neighbors. 
This allows us to vary the influence of a vertex's neighbors on its probability of activation between these two
subgroups. We suggest that in future analyses this may provide important insights into the role of group
structure and peer influence in processes of social contagion, such as opinion formation
\cite{friedkin:86,friedkin:11}.

%***************************************
\section{Cascade Analysis} \label{sec:2}
%***************************************

As was also the case for edge-triangle graphs \cite{hackett:11}, in order to extend to clique-based graphs the
approach of \cite{gleeson:08} we must first reconcile the presence of clustering with the locally treelike
approximation on which that approach is founded. In considering how best to proceed, let us return briefly to
\cite{gleeson:09a} and remind ourselves of the structural properties of the $\gm(k,c)$ ensemble.

\begin{figure}[t]
\centering
\includegraphics[width=\columnwidth]{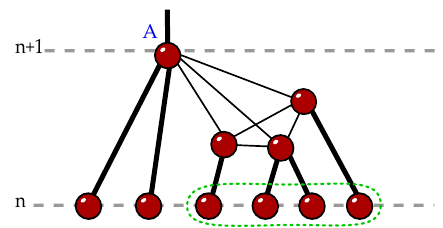}
\caption{(Color online) Level-by-level cascade propagation in a $\gm(k,c)$ graph using the tree approximation. External edges
emphasized.}
\label{fig:1}
\end{figure}

In Fig.~\ref{fig:1} we have reproduced Fig.~2 of \cite{gleeson:09a}. This figure shows a portion of an arbitrary
$\gm(k,c)$ graph that has been reconfigured into a treelike formation. The essential characteristics of this
reconfiguration can be explained most succinctly by looking at the local edge topology of the randomly chosen
vertex $A$. This vertex, positioned on level $n+1$ of the tree, has degree $k=6$ and is a member of a
$4$-clique. Its six incident edges are made up of $c-1=3$ internal edges, which connect $A$ to its neighboring
clique members, and $k-c+1=3$ external edges (emphasized). Of these external edges, one connects $A$ to its
parent vertex on the next level up, while the remaining $k-c=2$ connect $A$ to its external children on level
$n$. The clique neighbors are positioned on an unlabelled intermediate level between $A$ and its
grandchildren (circled with dashed line) on level $n$. This categorization and positioning of vertices is
representative of how the tree-based framework operates throughout the graph. Note that any vertex may be
treated similarly to $A$ regardless of the size of the clique to which it belongs. For any $(k,c)$ pairing such
that $k\geq c-1$ (see \cite{gleeson:09a}), $c-1$ clique neighbors can always be made to reside in the
interspace between a vertex and the level below and one may also stipulate in general that at most one
external edge leads to the parent above. In extreme cases, a vertex with no internal edges is simply a member of
a $1$-clique and therefore all of its connections will pass directly from one level to the next ($c-1=0$), as
in \cite{gleeson:07}. A vertex with no external edges must reside either at the root of the tree and have no
parent if it is part of a clique or it must be entirely isolated and have zero connections in total.

This, then, was the key that allowed Gleeson to calculate the giant connected component (GCC) size $S$ in bond
percolation on $\gm(k,c)$ graphs. Equation~(5) of \cite{gleeson:09a} was used to determine the conditional
probability that a vertex like $A$ is active (part of the GCC) on each level of the tree and Eq.~(6) of
\cite{gleeson:09a} then gave $S$ as the probability of activation of the root vertex by using the steady-state
value from Eq.~(5). The restriction of this theory to bond percolation arises primarily from its reliance on a
set of polynomials that were defined and tabulated by Newman in \cite{newman:03a}. Crucially, however, those
polynomials play no role in the conceptualization described above. Thus our task of extending the theory of
\cite{gleeson:08} amounts to taking this framework and introducing the response function mechanism. Since we
shall not apply the polynomials of \cite{newman:03a}, a straightforward substitution of $F_{m}^{k}$ will not
suffice. In fact, as we will now show, our approach requires a set of equations entirely different from those 
of \cite{gleeson:09a}.

%**************************************************
\subsection{Expected cascade size} \label{ssec:2.1}
%**************************************************

With the theoretical foundations in place, we can begin to derive generalized analytical expressions for 
cascades on $\gm(k,c)$ graphs. We proceed in the familiar manner by considering the probability $q_{n+1}$
that the randomly selected vertex $A$ in Fig.~\ref{fig:1} is active, conditional on its parent vertex being
inactive. As is usual for the tree-based approach, we stipulate that the vertex $A$ can become active only due
to the influence of the states of the neighboring vertices directly below it in the tree. In this case,
however, $A$ has two different types of neighbors: It has $k-c$ external children on level $n$ and $c-1$ clique
neighbors on the intermediate level. Significantly, the ways in which these two types of neighbor can become
active in their own right are quite distinct from each other. Thus their contributions to the probability of
activation of $A$ must be calculated separately. This is the first problem to be addressed.

Starting with the simpler of the two contributions, let us write down the probability that an arbitrary number,
call it $j$, of $A$'s external neighbors are active. Since there is no clustering between these vertices, each
one is independently activated by its own children on level $n-1$ with probability $q_{n}$. Therefore, the
probability that a total of $j$ out of $k-c$ external neighbors are activated in this way is given simply by
the binomial probability mass function (PMF)
\begin{equation}
B_{j}^{k-c}(q_{n})={k-c \choose j}{q_{n}}^{j}(1-q_{n})^{k-c-j}.
\label{eq:1}
\end{equation}

For the second contribution to $A$, matters are made considerably more complicated by the fact that its $c-1$
clique neighbors are fully connected. This means that the probability that each of these clique neighbors is
active depends not only on the states of their children---the four grandchildren of $A$ on level $n$---but also
on the states of one another. Recall from the derivation of our theory for cascades on $p(s,t)$ graphs in
\cite{hackett:11} that we had to account for the fact that each vertex at the base of a triangle can directly
influence the state of the other. We are faced with a similar problem here; however, since we are now dealing
with $\gm(k,c)$ graphs we have a whole spectrum of clique sizes to contend with.

One can appreciate how much more intricate this will make our calculations by imagining that $A$ were part of a
very large clique [as it could be, depending on our choice of $\gm(k,c)$]. For example, if $A$ were in a
$10$-clique, then $c-1=9$ intermediate vertices would each have a role to play in determining each others'
states. The solution in this case would require an extensive list of combinatorial expressions similar to, but
extending far beyond, Eqs.~(5)--(8) of \cite{hackett:11}. Ideally, we would like to avoid tabulating
combinatorial terms altogether and instead have a single compact analytical expression that is flexible enough
to deal with any clique size. This expression would allow us to feed in the total number of clique neighbors as
a variable and would then return the probability that a certain fraction of them are active. Evidently, the
derivation of such an expression is not straightforward. We shall therefore postpone this task until later in
our presentation.

In the meantime, we continue our analysis of cascade propagation by simply providing the name of this function
and taking it for granted that later in Sec.~\ref{sec:3} we will define precisely how it operates. Let us call
the relevant function $R_{m}^{c-1}(q_{n})$ and in doing so refer to it as the probability that in a clique of
$c-1$ intermediate vertices a total of $m$ are active, conditional on the top vertex of the $c$-clique to
which they belong (vertex $A$ in Fig.~\ref{fig:1}) being inactive. The dependence on $q_{n}$ arises from the
fact that each intermediate vertex has its own set of children on level $n$ and each of those children
($A$'s grandchildren in Fig.~\ref{fig:1}) is active with probability $q_{n}$. Summing over all possible
values of $m$ gives $\sum_{m=0}^{c-1}R_{m}^{c-1}(q_n)=1$.

If we accept the meaning of the label $R_{m}^{c-1}(q_{n})$ and combine it with Eq.~(\ref{eq:1}) above, we now
have the necessary terms in which to express the contribution of $A$'s external children and clique neighbors
towards its probability of activation $q_{n+1}$. This takes us very close to defining an iterative equation for
$q_{n+1}$ in terms of $q_{n}$. The missing ingredient is the probability $\zt(k,c)$ that the random vertex
$A$, while having degree $k$ and being a member of a $c$-clique, is also the child of a random vertex on
level $n+2$. This probability plays a role similar to that of the term $(k/z)p_{k}$ in Eq.~(1) of
\cite{gleeson:08}, which gives the probability of reaching a child of degree $k$ by traveling along a randomly
chosen edge from its parent in a nonclustered graph (see \cite{newman:03b}). Similarly, here $\zt(k,c)$ closes
our iteration by allowing us to average over all vertices on level $n+1$ in the correct manner. We express this
probability as

\begin{equation}
\zt(k,c)=(k-c+1)\gm(k,c)/z_{e},
\label{eq:2}
\end{equation}
where $z_{e}=\sum_{k,c}(k-c+1)\gm(k,c)$ is the average number of external edges per vertex.

Combining all three of our ingredients, we can now write our generalized iterative equation in terms of an
arbitrary response function $F_{m+j}^{k}$ as
\begin{equation}
q_{n+1}=\rho_{0}+(1-\rho_{0})\sum_{k,c}\zeta(k,c)\Psi(q_{n},k-1),
\label{eq:3}
\end{equation}
where
\begin{equation}
\Psi(q_{n},x)=\sum_{j=0}^{x-c+1}\sum_{m=0}^{c-1}B_{j}^{x-c+1}(q_{n})R_{m}^{c-1}(q_{n})F_{m+j}^{k}.
\label{eq:4}
\end{equation}
Thus we have derived an analytical expression for the probability that a randomly chosen vertex on the next
level up, generically called $n+1$, is active, conditional on its parent being inactive. Referring once again to
Fig.~\ref{fig:1}, Eq.~(\ref{eq:3}) tells us that the vertex $A$ will be found active if it was initially
activated as part of the seed fraction $\rho_{0}$ or (with probability $1-\rho_{0}$) if it subsequently became
active in response to the states of the $x=k-1$ neighbors directly below it in the tree. For the latter,
Eq.~(\ref{eq:4}) indicates that there are two distinct contributions from two different sets of neighbors: one
from the external children of $A$ and the other from the intermediate clique members. A total of $j$ of the
first type of neighbor are active with probability $B_{j}^{k-c} (q_{n})$ and $m$ of the second type with
probability $R_{m}^{c-1}(q_{n})$. Whether the sum of $j$ and $m$ is sufficient to activate $A$ is determined by
the response function $F_{m+j}^{k}$.

In the usual manner, iterating Eq.~(\ref{eq:3}) to the steady state will give us $q_{\infty}$. This value can
then be used in the following expression to determine the probability of activation of the root
vertex:
\begin{equation}
\rho=\rho_{0}+(1-\rho_{0})\sum_{k,c}\gamma(k,c)\Psi(q_{\infty},k).
\label{eq:5}
\end{equation}
The probability $\rho$ is equivalent to the expected cascade size (see the discussion in \cite{hackett:11}). The
differences between this equation and Eq.~(\ref{eq:3}) above are attributable to the fact that the root vertex
has no parent. This means that all of the root's $k$ edges extend downwards to its children, hence
$\Psi(q_{\infty},k)$. It also means that the correct term for averaging is simply $\gamma(k,c)$.

Taken together, then, Eqs.~(\ref{eq:3})-(\ref{eq:5}) constitute the core of our present analytical approach. We can
use these equations to investigate various different cascade processes by applying the appropriate definition of
the response function $F_{m+j}^{k}$ in each case. In Sec.~\ref{sec:4} we will provide the definitions of
$F_{m+j}^{k}$ for bond percolation and Watts's model. Before that, we must also define the function $R_{m}^{c-1}
(q_{n})$. This task will occupy all of Sec.~\ref{sec:3}. Next, let us conclude Sec.~\ref{sec:2} by deriving a
general first-order cascade condition.

%**********************************************
\subsection{Cascade condition} \label{ssec:2.2}
%**********************************************

The cascade condition determines whether an infinitesimally small seed fraction $\rho_{0}$ of active vertices
will generate a nonvanishing mean cascade size as the total number of vertices in the graph diverges
($N\rightarrow\infty$). For this to happen the iteration of Eq.~(\ref{eq:3}) must cause the activation
probability $q_{n}$ to grow from an initial value $q_{0}=0$ to a nonzero steady-state $q_{\infty}$
\cite{gleeson:08}. If we regard Eq.~(\ref{eq:3}) (with $\rho_0=0$) as a nonlinear function of $q$ of the general
form $q_{n+1}=H(q_n)$, then this last condition can be expressed, to first-order, as $H'(0)>1$.

To evaluate $H'(0)$ we require the following results for the binomial PMF of Eq.~(\ref{eq:1}):
\begin{equation}
B_{j}^{k-c}(0)=\dl_{j,0},
\label{eq:6}
\end{equation}
\begin{equation}
\frac{d}{dq}B_{j}^{k-c}(q)\Bigg|_{q=0}=(k-c)(\dl_{j,1}-\dl_{j,0}).
\label{eq:7}
\end{equation}
Using Eqs.~(\ref{eq:6}) and (\ref{eq:7}) in Eq.~(\ref{eq:3}), we find that the first derivative of $H(q)$,
evaluated at $q=0$, may be expressed as
\begin{align}
\nn H'(0)=\sum_{k,c}\zeta(k,c)\sum_{m=0}^{c-1}\Bigg[(k-c)\big(F_{m+1}^{k}-F_{m}^{k}\big)&\\
\times R_{m}^{c-1}(0)+F_{m}^{k}\frac{d}{dq}R_{m}^{c-1}(q)\Bigg|_{q=0}&\Bigg].
\label{eq:8}
\end{align}
This is the left-hand side of our cascade condition. Note that because this expression depends on $R_{m}^{c-1}
(0)$ and the first derivative of $R_{m}^{c-1}(q)$ at $q=0$ it becomes an increasingly arduous task to calculate
$H'(0)$ from Eq.~(\ref{eq:8}) as the size of the largest clique in our graph increases. As we shall see in the
next section, the evaluation of the function $R_{m}^{c-1}(q)$ becomes increasingly difficult as the value of $c$
increases. For this reason, in our analysis in Sec.~\ref{sec:4} we will choose $\gm(k,c)$ such that the cliques
in our graphs are constrained to sizes of $c\leq 4$. In addition, we shall make the simplifying assumption that
$F_{0}^{k}=0$ (see \cite{hackett:11}). This implies that a vertex will never activate if none of its neighbors
are active and is a suitable approximation for the calculation of our first-order condition.

%**********************************************
\section{Active Clique Neighbors} \label{sec:3}
%**********************************************

Backtracking slightly in the flow of our presentation, we will now derive a concise closed-form expression for
the probability labelled above as $R_{m}^{c-1}(q_{n})$. Let us begin by recapitulating the meaning of
this label. According to our earlier definition, it is the probability that $m$ out of $c-1$ intermediate level
$c$-clique vertices are active, given that their own externally linked children are each independently active
with probability $q_n$ and that the parent vertex at the top of the $c$-clique is inactive. In Fig.~\ref{fig:1},
for example, $R_{m}^{3}(q_{n})$ is the probability that $m$ of the vertex $A$'s three clique neighbors are active,
given that each of the four grandchildren of $A$ (circled) has an activation probability of $q_n$ and that $A$
is itself inactive.

In considering how to calculate $R_{m}^{c-1}(q_{n})$ in general, we see immediately that it is not the states of
the external grandchildren that will cause us difficulty, but rather the fact that the state of each
intermediate clique member can influence the states of all other members. In our framework, every $c$-clique has
one of its (internally linked) members designated as the parent and placed on level $n+1$. This leaves each of
the remaining $c-1$ clique members on the intermediate level with $k-c+1$ external edges to connect to its own
children on level $n$. The probability that some number $j$ of these children are active is given by the
binomial PMF $B_{j}^{k-c+1}(q_{n})$. Thus the probability that an intermediate clique member is activated by
its children is quite easy to calculate. In contrast, in order to deal with the influence of the $c-1$
clique members on one another, we will have to consider carefully the various combinations of states that may
exist within the intermediate portion of the clique.

Our first step in tackling this problem is to provide a mechanism for the intermediate clique members to be
activated, which combines both internal and external influences. We define
\begin{equation}
G_{d}^{c-2}(q_{n})=\sum_{k}\frac{\gamma(k,c)}{p_{c}}\sum_{j=0}^{k-c+1}B_{j}^{k-c+1}(q_{n})F_{d+j}^{k},
\label{eq:9}
\end{equation}
for $c\geq2$, as the conditional probability that an intermediate $c$-clique vertex will be activated if $d$ of
its $c-2$ clique neighbors on the same level are active, given its external children are each active with
probability $q_{n}$ and its parent on level $n+1$ is inactive. The term $\gm(k,c)/p_{c}$ is the degree 
distribution of vertices that belong to a $c$-clique, where $p_{c}=\sum_{k}\gm(k,c)$. The response
function $F_{d+j}^{k}$ will determine whether $d$ active neighbors plus $j$ active children are enough to cause
activation. Defined as such, $G_{d}^{c-2}(q_{n})$ provides a fundamental term in which to express the various
possible active configurations, thus permitting us to begin the procedure of counting.

\begin{figure}[t]
\centering
\includegraphics[width=\columnwidth]{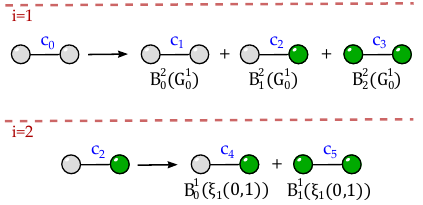}
\caption{(Color online) Transition probabilities for a pair ($c-1=2$) of intermediate clique neighbors in a $\gm(k,c)$ graph.
Colour indicates vertex state: light gray, inactive; dark gray (green), active.}
\label{fig:2}
\end{figure}

We consider first the simplest nontrivial case, namely, $c=3$. Suppose we pick from some arbitrary $\gm(k,c)$
graph a vertex with degree $k$ that is also a member of a $3$-clique. If we let this vertex reside on level
$n+1$ of the tree and also position its $c-1=2$ clique neighbors between level $n+1$ and level $n$ below, our
task then is to calculate $R_{m}^{2}(q_{n})$. To do this, let us refer to Fig.~\ref{fig:2} and look at the
possible states of these two vertices in isolation from their inactive parent.

Starting with both vertices inactive---the configuration labelled $c_{0}$ in Fig.~\ref{fig:2}---we first count
the possible configurations of states after one round ($i=1$) of synchronous updates. Since we have started from
$c_{0}$, with both vertices inactive, the probability of either vertex becoming active in this first round is
simply $G_{0}^{1}$. Therefore, each possible outcome ($c_{1}$, $c_{2}$, or $c_{3}$ in Fig.~\ref{fig:2}) is
determined by a binomial PMF with probability of success $G_{0}^{1}$. Configuration $c_{1}$, in which both
vertices have remained inactive, will occur with probability $B_{0}^{2}(G_{0}^{1})$. Similarly, configuration
$c_{2}$, in which one vertex has been activated and the other has remained inactive, will occur with probability
$B_{1}^{2}(G_{0}^{1})$. Finally, configuration $c_{3}$, in which both vertices have been activated, will occur
with probability $B_{2}^{2}(G_{0}^{1})$. [Note that in each term $G_{d}^{c-2}\equiv G_{d}^{c-2}(q_{n})$; we will
use this abbreviation throughout.]

Having determined the three distinct outcomes of the first round of updates, we will now categorize each
configuration into either of two types: terminal or volatile. In a terminal configuration no
further changes of state are possible because all vertices have reached their own steady-state of either
permanent activation or inactivation. In a volatile configuration, however, there exists at least one
inactive vertex that is liable to become active. Thus, as long as volatile configurations are produced we must
continue with another round of updates. The process of updating will reach its end when all configurations are
terminal. Categorizing the outcomes of round one tells us whether or not a second round is necessary and also
indicates which configurations need to be updated. Configuration $c_{1}$ is clearly terminal since the
transition from $c_{0}$ to $c_{1}$ has established that neither vertex can activate while the other remains
inactive. Similarly, $c_{3}$ is also terminal for the simple reason that we do not allow active vertices to
revert to being inactive. Configuration $c_{2}$, however, is volatile since the transition from $c_{0}$ to
$c_{2}$ has shown us that one of these vertices can activate without the other first being active, but that the
same is not true of this other vertex. That is to say, we know that the inactive vertex in $c_{2}$ cannot
activate without an active neighbor. What is not clear from $c_{2}$ is whether the vertex that did activate in
round one is now sufficient to activate the vertex that remained inactive in that round. The only way to
determine this is to run a second round ($i=2$) of updates on $c_{2}$.

As was the case in the first round, to begin the second round we must provide an appropriate probability of
activation. We want to know if the active vertex in $c_{2}$ is enough to activate the inactive vertex in
$c_{2}$, given that the inactive vertex cannot activate without an active neighbor. This can be decided upon by
using the activation probability $\xi_{1}(0,1)$ defined by the function
\begin{equation}
\xi_{c-2}(a,b)=\frac{G_{b}^{c-2}(q_{n})-G_{a}^{c-2}(q_{n})}{1-G_{a}^{c-2}(q_{n})}.
\label{eq:10}
\end{equation}
Equation~(\ref{eq:10}) gives us the conditional probability that in a clique of $c-1$ intermediate vertices
$b$ active vertices are enough to cause the activation of one of their inactive clique neighbors, given that
$a$ active vertices are insufficient to do so. The function $\xi_{c-2}(a,b)$ is defined for $0\leq a\leq b\leq c-2$
and is non-negative for all such values since by Eq.~(\ref{eq:9}) $G_{d}^{c-2}(q_n)$ is an increasing function
of $d$. This latter property is true of $G_{d}^{c-2}(q_n)$ since $F_{m}^{k}$ is defined (see Sec.~\ref{sec:1}) to
be a nondecreasing function of $m$ [and therefore so is $F_{d+j}^k$ in Eq.~(\ref{eq:9})]. The configurations
produced by updating with this probability are once again given by a binomial PMF. With probability
$B_{0}^{1}(\xi_{1}(0,1))$ the inactive vertex will remain inactive, thereby producing configuration
$c_{4}$. Conversely, with probability $B_{1}^{1}(\xi_{1}(0,1))$ the inactive vertex will activate, thereby
producing configuration $c_{5}$. Categorizing $c_{4}$ and $c_{5}$, we find both configurations are terminal 
and therefore the process of updating may now cease.

With all terminal configurations now achieved, the next step in our derivation of $R_{m}^{2}(q_{n})$ is to
combine the various transition probabilities listed in Fig.~\ref{fig:2} and use them to calculate each of
$R_{0}^{2}(q_{n})$, $R_{1}^{2}(q_{n})$, and $R_{2}^{2}(q_{n})$. Tracing our way through Fig.~\ref{fig:2}, we
reach a terminal state in which no vertices are active by following the route $c_{0}\rightarrow c_{1}$.
Similarly, we end with one active vertex by following $c_{0}\rightarrow c_{2}\rightarrow c_{4}$. Finally, a
terminal state with two active vertices is given by either of the routes $c_{0}\rightarrow c_{3}$ or
$c_{0}\rightarrow c_{2}\rightarrow c_{5}$. All of this information can be expressed succinctly using the various
transition probabilities associated with each route if we bear in mind that a transition from one configuration
to another, symbolized by $\rightarrow$, corresponds to the multiplication of probabilities, and also that the
word \emph{or} corresponds to addition. To summarize, the set of routes described here yields the following set
of equations:
\begin{equation}
R_{0}^{2}(q_{n})=B_{0}^{2}(G_{0}^{1}),
\label{eq:11}
\end{equation}
%\vspace{-6mm}
\begin{equation}
R_{1}^{2}(q_{n})=B_{1}^{2}(G_{0}^{1})B_{0}^{1}(\xi_{1}(0,1)),
\label{eq:12}
\end{equation}
%\vspace{-6mm}
\begin{equation}
R_{2}^{2}(q_{n})=B_{1}^{2}(G_{0}^{1})B_{1}^{1}(\xi_{1}(0,1))+B_{2}^{2}(G_{0}^{1}).
\label{eq:13}
\end{equation}

The final step towards our goal of writing a closed-form expression for $R_{m}^{2}(q_{n})$ is to
find a way of expressing Eqs.~(\ref{eq:11})--(\ref{eq:13}) as the outputs of a single function that has been
given the inputs $m = 0$, $1$, and $2$, respectively. There may be a number of different ways of defining such
a function, some of which may appear more elegant than others. For our own part, we can offer a particularly
concise definition by introducing a new variable and considering how the various combinations of states
determined by Eqs.~(\ref{eq:11})--(\ref{eq:13}) can be reproduced in a parsimonious manner.

Our new variable is called $l_{i}$. We define it as the number of new activations in round $i$ of synchronous
updates. In the scheme presented above we had two rounds; therefore, we define the pair $l=(l_{1}, l_{2})$ as
the sequence of new activations over both rounds. This allows us to represent all possible routes through the
configurations of Fig.~\ref{fig:2} as a collection of ordered pairs. For example, $l=(1,0)$ means that there
is one activation in round $i=1$ and no activations in round $i=2$ and therefore corresponds to the route
$c_{0}\rightarrow c_{2}\rightarrow c_{4}$. Similarly, $l=(1,1)$ corresponds to
$c_{0}\rightarrow c_{2}\rightarrow c_{5}$. By applying this notation we find that the following
equation will reproduce each of the Eqs.~(\ref{eq:11})--(\ref{eq:13}) above:
\begin{align}
R_{m}^{2}(q_{n})=\sum_{l_{1}+l_{2}=m}B_{l_1}^{2}\left(G_{0}^1\right)B_{l_2}^{2-l_1}\left(\xi_{1}(0,l_1)\right).
\label{eq:14}
\end{align}
Note that the summation $\sum_{l_{1}+l_{2}=m}$ in Eq.~(\ref{eq:14}) is taken over all pairs $l=(l_{1},l_{2})$
such that $l_{1}+l_{2}=m$, where $m$ is the total number of active vertices.

To demonstrate how Eq.~(\ref{eq:14}) operates let us calculate $R_{1}^{2}(q_{n})$ by setting $m=1$. The set of
all $l$ pairs that add up to this value of $m$ is $l\in\{(0,1),(1,0)\}$. Substituting each of these pairs in
turn into the right hand side of Eq.~(\ref{eq:14}) and then summing gives
$R_{1}^{2}(q_{n})=\big[0+2G_{0}^{1}\big(1-G_{1}^{1}\big)\big]$, thereby reproducing Eq.~(\ref{eq:12}) above. The
values of $R_{0}^{2}(q_{n})$ and $R_{2}^{2}(q_{n})$ are found similarly by using the parameters $m=0$ and
$l=(0,0)$, and $m=2$ and $l\in\{(0,2),(1,1),(2,0)\}$, respectively.

Thus, in Eq.~(\ref{eq:14}) we have found an expression for $R_{m}^{2}(q_{n})$, which, we remind ourselves once
more, is the conditional probability that $m$ of the two intermediate vertices in a $3$-clique are active, given
that each of their own children are active with probability $q_{n}$, and that the vertex at the top of the clique is
inactive. Recall, however, that our ultimate goal is to provide a general expression for $R_{m}^{c-1}(q_{n})$.
Our approach to this problem has been to determine a series of expressions for increasing values of $c$ and then
to express each of these as special cases of a single unifying expression. Each individual expression for
$R_{m}^{c-1}(q_{n})$, where $c>3$, can be found by a method similar to the one described above for $R_{m}^{2}(q_{n})$.
The core of this method is the same regardless of the value of $c$ and can be summarized in general as follows.
\begin{enumerate}[(i)]
  \item Simultaneously update the states of all inactive vertices.
  \item Categorize the resulting configurations of states as either terminal or volatile, removing those that
  are terminal from further consideration.
  \item Repeat steps (i) and (ii) until no volatile configurations remain.
\end{enumerate}

Counting the terminal configurations will then provide the various outcomes obtainable in the steady state of
the cascade. For example, in determining $R_{m}^{3}(q_{n})$, the application of these three steps reveals every
possible active configuration in a triangle of connected vertices and each associated transition probability.
As above, following the different routes towards each terminal configuration indicates the correct sequence of
multiplications and additions to employ to calculate the values of $R_{m}^{3}(q_{n})$ for $0\leq m\leq 3$. This
procedure yields the following set of equations:
\begin{equation}
R_{0}^{3}(q_{n})=B_{0}^{3}(G_{0}^{2}),
\label{eq:15}
\end{equation}
\vspace{-6mm}
\begin{equation}
R_{1}^{3}(q_{n})=B_{1}^{3}(G_{0}^{2})B_{0}^{2}(\xi_{2}(0,1)),
\label{eq:16}
\end{equation}
\vspace{-6mm}
\begin{align}
\nn R_{2}^{3}(q_{n})=&B_{2}^{3}(G_{0}^{2}) B_{1}^{2}(\xi_{2}(0,1))B_{0}^{1}(\xi_{2}(1,2))\\
&+B_{2}^{3}(G_{0}^{2})B_{0}^{1}(\xi_{2}(0,2)),
\label{eq:17}
\end{align}
\vspace{-6mm}
\begin{align}
\nn R_{3}^{3}(q_{n})=&B_{1}^{3}(G_{0}^{2})B_{1}^{2}(\xi_{2}(0,1))B_{1}^{1}(\xi_{2}(1,2))\\
\nn &+B_{3}^{3}(G_{0}^{2})+B_{2}^{3}(G_{0}^{2})B_{1}^{1}(\xi_{2}(0,2))\\
&+B_{1}^{3}(G_{0}^{2})B_{2}^{2}(\xi_{2}(0,1)).
\label{eq:18}
\end{align}

Continuing in the same manner as before, an expression for $R_{m}^{3}(q_{n})$ that contains
Eqs.~(\ref{eq:15})--(\ref{eq:18}) as special cases can be defined by applying the variable $l_{i}$ and
considering each unique sequence of activations $l=(l_{1},l_{2},l_{3})$. By doing this we have found that the
equation
\begin{align}
\nn R_{m}^{3}(q_{n})=\sum_{l_{1}+l_{2}+l_{3}=m}&B_{l_1}^{3}\left(G_{0}^2\right)B_{l_2}^{3-l_1}\left(\xi_{2}(0,l_1)\right)\\
&\times B_{l_3}^{3-(l_1+l_2)}\left(\xi_{2}(l_1,l_1+l_2)\right)
\label{eq:19}
\end{align}
will reproduce Eqs.~(\ref{eq:15})--(\ref{eq:18}).

Observe the similarities between equation Eq.~(\ref{eq:19}) and (\ref{eq:14}). They indicate that to create
an expression for $R_{m}^{3}(q_{n})$ from that for $R_{m}^{2}(q_{n})$ above all one must do (besides set
$c=4$) is place additional indices $l_{2}$ and $l_{3}$ in the appropriate positions and include one more
multiplicative term, namely, $B_{l_3}^{3-(l_1+l_2)}\left(\xi_{2}(l_1,l_1+l_2)\right)$. By running the entire scheme
of categorization and route counting over again with $c=5$ and $l=(l_{1},l_{2},l_{3},l_{4})$, we have observed
(in calculations not provided here) that a similar relationship also holds between $R_{m}^{3}(q_{n})$ and
$R_{m}^{4}(q_{n})$. The pattern of similarities detected in our calculations strongly suggests the following
form for a general expression for $R_{m}^{v}(q_{n})$, where $v$ is an integer $v\geq m$:
\begin{align}
R_{m}^{v}(q_{n})=&\sum_{|l|=m}\prod_{i=1}^{v} B_{l_i}^{n_{v,i}}(\theta_{v,i}).
\label{eq:20}
\end{align}

Let us unpack this expression. First, note that the variable $n_{v,i}$ in Eq.~(\ref{eq:20}) is defined
as $n_{v,i}=v-\sum_{j=1}^{i-1}l_{j}$ for $i\geq 2$ with $n_{v,1}=v$. Next, the variable
$\theta_{v,i}$ is defined as $\theta_{v,i}=\xi_{v-1}\left(\sum_{j=1}^{i-2}l_{j}, \sum_{j=1}^{i-1}l_{j}\right)$
for $i\geq 3$ with $\theta_{v,1}=G_{0}^{v-1}$ and $\theta_{v,2}=\xi_{v-1}(0,l_{1})$. Finally, the term $|l|$ in
the summation of Eq.~(\ref{eq:20}) is defined in multi-index notation (see, for example, \cite{wong:99}) as
$|l|=l_{1}+\ldots+l_{v}$.

By setting $v=c-1$ in Eq.~(\ref{eq:20}), we have the probability $R_{m}^{c-1}(q_n)$ expressed in closed form \footnote{The
condition $\sum_{m=0}^{c-1}R_{m}^{c-1}(q_{n})=1$ can be verified by a simple algebraic manipulation of the summation.}.
Applying this definition in Eqs.~(\ref{eq:3})-(\ref{eq:5}) (see Sec.~\ref{sec:2}) completes our analytical description
of cascades on clique-based graphs and permits us to proceed with the task of verifying our approach. We will provide
this verification in the next section by comparing predicted values of the expected cascade size from Eq.~(\ref{eq:5})
against the results of numerical simulations of bond percolation and Watts's model.

It must be noted, however, that as the size of the largest clique of in our graph $c_{\textrm{max}}$ increases it becomes
more and more computationally intensive to evaluate $R_{m}^{c-1}(q_{n})$ using Eq.~(\ref{eq:20}). This is
primarily because of the exponentially increasing number of possible combinations for the multi-index $l$ as the
number of active clique members to be counted $m$ increases. It can be shown that the number of different
choices of $l$ that give nonzero contributions to the sum in Eq.~(\ref{eq:20}) is $2^{m-1}$.

%**********************************
\section{Simulations} \label{sec:4}
%**********************************

To test the theory of the previous two sections we require an appropriate set of definitions for the response
function $F_{m+j}^{k}$, corresponding to the processes in our familiar broad class (see
Sec.\hspace{0.5em}\ref{sec:1}). The function $F$, however, is the same one that has been used throughout our
groups' previous publications \cite{gleeson:08,gleeson:09a,hackett:11}. Gleeson began in \cite{gleeson:08} by
writing it in its simplest generalized form: $F_{m}^{k}$. There, it defined the probability that a $k$-degree
vertex in a locally tree-like graph may be activated by $m$ active neighbors. In \cite{hackett:11},
$F_{m}^{s+2t}$ gave the probability that a $k$-degree vertex in an edge-triangle graph may be activated by $m$
active neighbors, where $k=s+2t$. In the current presentation, $F_{m+j}^{k}$ prescribes the probability that a
$k$-degree vertex in a clique-based graph may be activated by $m+j$ active neighbors, where $j$ and $m$ are the
numbers of external and internal neighbors, respectively. Since $F$ has not changed (only its arguments have),
the same justifications of our use of the response function mechanism as were given in \cite{hackett:11} apply
equally here. Therefore, similarly to \cite{hackett:11}, the definitions of $F_{m+j}^{k}$ for different
processes are found by replacing $m$ with $m+j$ in the definitions of $F_{m}^{k}$ given in \cite{gleeson:08}.
With this aspect clarified, we can begin testing our approach against numerical simulations of various
processes.

%*********************************************
\subsection{Bond percolation} \label{ssec:4.1}
%*********************************************

We consider first the process of uniform bond percolation.
In this process each edge of the graph (external or internal) 
is deleted with probability $1-\phi_b$. The quantity $\phi_b$ is 
the bond occupation probability and nondamaged edges are termed 
occupied. Replacing $m$ with $m+j$ in Eq.~(6) of \cite{gleeson:08} 
defines $F_{m+j}^{k}$ for this process:
\begin{equation}
F_{m+j}^{k}=1-(1-\phi_b)^{m+j}.
\label{eq:21}
\end{equation}
Applying this definition in the respective $\rho_{0}\rightarrow0$ 
limits of Eqs.~(\ref{eq:3})-(\ref{eq:5}) above allows us to use these 
equations to calculate the expected GCC size $S$ of a clique-based 
graph, which is nonzero for $\phi_b>\widehat{\phi_{b}}$. This critical 
value, $\widehat{\phi_{b}}$, is known as the bond percolation threshold.

%We consider first the process of uniform bond percolation. Replacing $m$ with $m+j$ in Eq.~(6) of
%\cite{gleeson:08} defines $F_{m+j}^{k}$ for this process:
%\begin{equation}
%F_{m+j}^{k}=1-(1-\phi_b)^{m+j}.
%\label{eq:22}
%\end{equation}
%Applying this definition in the respective $\rho_{0}\rightarrow0$ limits of Eqs.~(\ref{eq:3})-(\ref{eq:5}) above
%allows us to use these equations to calculate the expected GCC size, $S$.

\begin{figure}[t]
\centering
\includegraphics[width=\columnwidth]{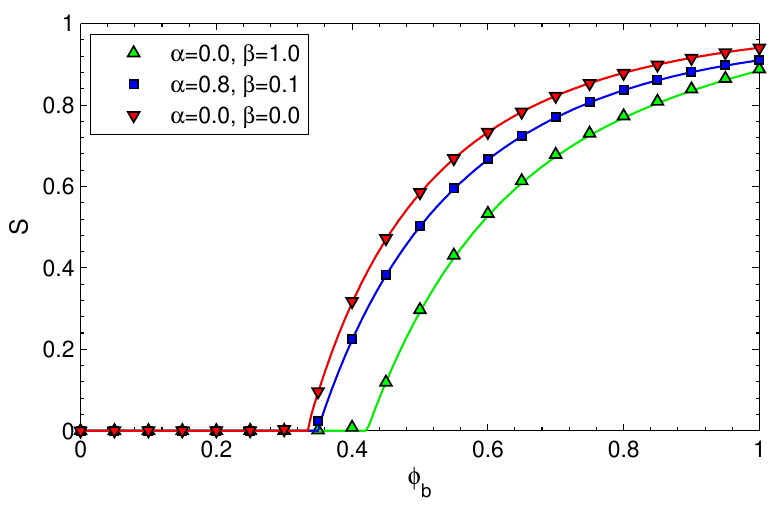}
\caption{(Color online) Bond percolation on $\gm(k,c)$ graphs of $N=10^5$ vertices and Poisson degree distribution $p_{k}$
with mean degree $z=3$. Numerical simulations (symbols) averaged over $100$ realizations and theory of
Sec.~\ref{sec:2} (lines) on a plot of GCC size $S$ vs. bond occupation probability $\phi_{b}$.}
\label{fig:3}
\end{figure}

In Fig.~\ref{fig:3} we have plotted our calculations of $S$ from Eq.~(\ref{eq:5}) against the results of
numerically simulated $\gm(k,c)$ graphs (see the caption). The parameters chosen for this figure are the same as
those used in Fig.~3(a) of \cite{gleeson:09a}. Each graph has a Poisson degree distribution
$p_{k}=z^{k}e^{-z}/k!$ with mean degree $z=3$. Following \cite{gleeson:09a}, we set
$\gm(k,c)=\big[(1-\al-\bt)\dl_{c,1}+\al\dl_{c,3}+\bt\dl_{c,4}\big]p_{k}$ for $k\geq3$, where $\al,\bt\in[0,1]$.
In this way we create nonzero clustering by assigning a fraction $\al$ of $k$-degree vertices to $3$-cliques and
a fraction $\bt$ to $4$-cliques. Additionally, since a $2$-degree vertex cannot belong to a clique of size
$c>3$, we assign a fraction $\al$ of these vertices to $3$-cliques using
$\gm(2,c)=\big[(1-\al)\dl_{c,1}+\al\dl_{c,3}\big]p_{2}$. We let vertices of degree zero or one belong to
$1$-cliques: $\gm(k,c)=p_{k}\dl_{c,1}$. This choice of $\gm(k,c)$ limits the largest clique size to $c_{\textrm{max}}=4$
and therefore makes the evaluation of $R_{m}^{c-1}(q_{n})$ relatively simple. By varying $\al$ and $\bt$
different levels of clustering can be prescribed. Again following \cite{gleeson:09a}, we use three ($\al,\bt$)
pairs: ($0,0$), ($0.8,0.1$), and ($0,1$). Evidently, ($0,0$) produces a nonclustered graph (downward-pointing triangles). 
We can use Eq.~(2) of \cite{gleeson:09a} to define the global clustering coefficient $C_{2}=\sum_{k}p_{k}c_{k}$. From this
one may show that ($0.8,0.1$) produces a clustered graph with $C_{2}=0.31$ (squares), and also that ($0,1$) gives a
graph with $C_{2}=0.35$ (upward-pointing triangles).

The percolation thresholds for each nonzero value of $C_{2}$ can be calculated from our cascade condition of
Sec.~\ref{sec:2} by setting $H'(0)=1$ in Eq.~(\ref{eq:8}) and solving  for $\phi_{b}$ (see \cite{hackett:11}).
This of course requires that we first substitute Eq.~(\ref{eq:21}) into Eq.~(\ref{eq:8}). We also require the
following results for the function $G_{d}^{c-2}(q)$ of Eq.~(\ref{eq:9}) in order to evaluate $R_{m}^{c-1}(0)$
and the first derivative of $R_{m}^{c-1}(q)$ at $q=0$:

\begin{equation}
G_{d}^{c-2}(0)=\sum_{k}\frac{\gm(k,c)}{p_c}F_{d}^{k},
\label{eq:22}
\end{equation}
\begin{equation}
\frac{d}{dq}G_{d}^{c-2}(q)\Bigg|_{q=0}=\sum_{k}\frac{\gm(k,c)}{p_c}(k-c+1)\big(F_{d+1}^{k}-F_{d}^{k}\big).
\label{eq:23}
\end{equation}

Using Eqs.~(\ref{eq:21})-(\ref{eq:23}) in Eq.~(\ref{eq:8}) we calculate the threshold for $C_{2}=0.31$ to be
$\wh{\phi_{b}}=0.349$, while for $C_{2}=0.35$ we get $\wh{\phi_{b}}=0.423$. The threshold for $C_{2}=0$ is
simply the configuration model value $\wh{\phi_{b}}=1/z$ \cite{callaway:00}.

The match obtained between theory and numerics in Fig.~\ref{fig:3} provides a clear validation of our approach
in the case of bond percolation. Furthermore, because we have chosen the same parameters as Fig.~3(a) of
\cite{gleeson:09a}, the results shown in that figure should correspond exactly with the results shown here in
Fig.~\ref{fig:3}. Comparing these two figures will reveal to the reader that they do indeed match. This
illustrates that our approach contains within its scope the ability to produce the same predicted values of $S$
as the theory of \cite{gleeson:09a}. However, as noted earlier at the beginning of Sec.~\ref{sec:2}, Gleeson's
equations depend on a set of polynomial functions defined and tabulated in \cite{newman:03a}. These polynomials
limit the application of his equations to bond percolation. The advantage of our approach over that of
\cite{gleeson:09a} is its purported applicability to other processes besides bond percolation. To
confirm that it really does possess this flexibility we consider for our second test Watts's model
\cite{watts:02}.

%******************************************
\subsection{Watts's model} \label{ssec:4.2}
%******************************************

Watts's model provides a simplified description of threshold-dependent cascade dynamics on complex networks. In
a sociological setting this model may provide a crude approximation of the processes of contagion that underlie
such phenomena as fashions, rumours, or popular opinions. Given, for example, a network of acquaintanceships
between a group of people, we can use Watts's model to calculate the steady-state fraction of active vertices in
the following binary-state decision process.

We begin by assigning a threshold $r_i$ drawn from the probability distribution $q(r)$ to each vertex
$1\leq i\leq N$ in the network. At each discrete time step $t$ the state of vertex $i$ is $v_i(t)\in[0,1]$,
where $v_i(t)=1$ indicates the participation of $i$ in the cascade and $v_i(t)=0$ indicates nonparticipation.
The dynamics is instigated by activating a small seed fraction of vertices at $t=0$. From $t=1$ until the
steady-state $\bar{t}$ the state of each vertex is updated synchronously at each $t$ according to the rule
\begin{align}
v_{i}(t)&=\begin{cases}
          \hspace{9.5mm} 1 \hspace{9.2mm} \text{if} \quad \frac{1}{k_{i}}\sum_{j}a_{ij}v_{j}(t)>r_{i}\\
          \hspace{1mm} \text{unchanged} \hspace{2mm} \text{otherwise},
          \end{cases}
\label{eq:24}
\end{align}
where $a_{ij}$ is the value in position $(i,j)$ of the adjacency matrix of the network and $k_i$ is the
degree of vertex $i$. By this mechanism vertex $i$ will join the cascade if the fraction of its direct
neighbors that are active exceeds its threshold, otherwise it will remain inactive. Once active, $i$ will
remain in this state.

In the steady-state the final fraction of active vertices is given by $\frac{1}{N}\sum_{i}v_i(\bar{t})$.
By averaging this last value over many individual runs of the model we can determine a numerical evaluation of
the expected cascade size $\rho$.

\begin{figure}[t]
\centering
\includegraphics[width=\columnwidth]{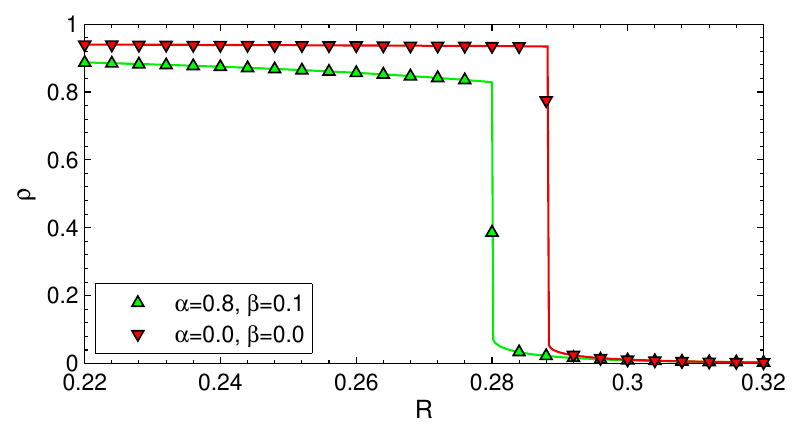}
\end{figure}
\begin{figure}[t]
\centering
\includegraphics[width=\columnwidth]{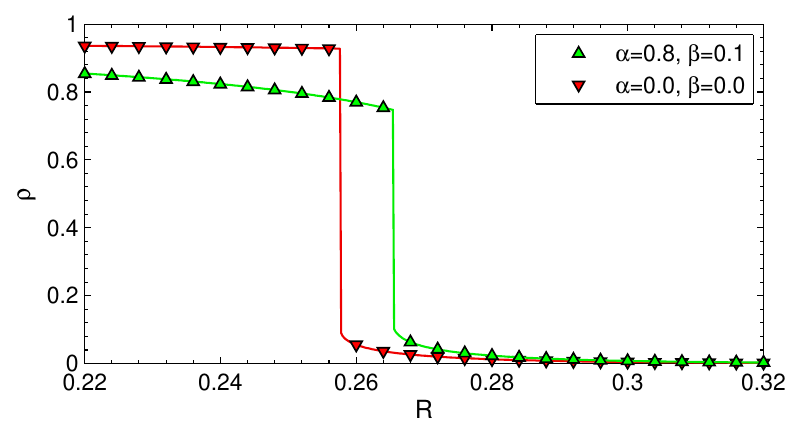}
\caption{(Color online) Watts's model on $\gm(k,c)$ graphs of $N=10^6$ vertices and Poisson degree distribution 
$p_{k}$ with mean degree $z=3$. Thresholds are drawn from a Gaussian distribution with mean $R$ and standard 
deviation $\sg=0.1$. Numerical simulations (symbols) averaged over $100$ realizations and theory of Sec.~\ref{sec:2} 
(lines) on a plot of cascade size $\rho$ vs. $R$. In (a) $w_{i}=1$ and $w_{e}=1$. In (b) $w_{i}=1.3$ and $w_{e}=0.85$.}
\label{fig:4}
\end{figure}

With the appropriate choice of response function $F_{m+j}^{k}$, our Eqs.~(\ref{eq:3})-(\ref{eq:5}) provide an
analytical match to the numerical results of Watts's model. In Fig.~\ref{fig:4} we present values of $\rho$
from Eq.~(\ref{eq:5}) plotted against the results of simulations on $\gm(k,c)$ graphs. The thresholds in each of
these graphs are drawn from a Gaussian distribution: $q(r)=N(R,0.1)$ (see the caption). Therefore, the response
function for our equations is defined by replacing $m$ with $m+j$ in Eq.~(2) of \cite{gleeson:08}:

\begin{equation}
F_{m+j}^{k}=\frac{1}{2}\Bigg[1+\textrm{erf}\Bigg(\frac{(m+j)/k-R}{\sg\sqrt{2}}\Bigg)\Bigg].
\label{eq:25}
\end{equation}
The choice of $q(r)=N(R,0.1)$ means some vertices will be assigned negative thresholds and will 
therefore automatically activate. This allows us to set $\rho_0=0$ in Eqs.~(\ref{eq:3})-(\ref{eq:5}). 
The structural variables used for this figure are the same as those applied previously in Fig.~\ref{fig:3}. All
graphs have Poisson degree distribution $p_{k}$ with $z=3$ and $\gm(k,c)$ is defined by the same three 
equations as above. We apply two $(\al,\bt)$ pairs, $(0,0)$ and $(0.8,0.1)$, corresponding to $C_2=0$ and 
$C_2=0.31$, respectively.

Figure~\ref{fig:4}{\color{blue}(a)} provides a further validation of our approach and explicitly demonstrates
its flexibility. In Fig.~\ref{fig:4}{\color{blue}(b)} we investigate a minor modification to Watts's model. The presence of
neighbors of two distinct kinds (internal and external) in clique-based graphs opens up some interesting
possibilities for the augmentation of the updating process described by Eq.~(\ref{eq:24}). To take one simple
example, consider the following weighting scheme. Let each active internal vertex have weight $w_i\in(0,\infty)$
and each active external vertex have weight $w_e\in(0,\infty)$. The response function for this process is given
by multiplying $m$ by $w_i$ and $j$ by $w_e$ in Eq.~(\ref{eq:25}). We propose that such a weighting may provide
insights into the role of group structure in determining the outcome of processes of social contagion such as 
those mentioned above. Problems of this nature have been of interest for quite some time 
(see \cite{friedkin:11} and references therein).

When $w_i=w_e=1$ we have the conventional version of Watts's model in which there is no bias
in favor of either type of neighbor [Fig.~\ref{fig:4}{\color{blue}(a)}]. However, settings where $w_i>w_e$ or
$w_i<w_e$ indicate a respective bias in favor of or against one's clique neighbors over one's external
neighbors. If we take the clique as a proxy for a tightly-knit social group, then the first setting describes a
scenario where the influence of ones peers is favored over influences from outside the immediate
peer group. The second setting describes the opposite scenario.

Figure~\ref{fig:4}{\color{blue}(b)} demonstrates why this modification of Watts's model is
interesting from an analytical perspective. Here we have set $w_i=1.3$ and $w_e=0.85$. Comparing this figure to
Fig.~\ref{fig:4}{\color{blue}(a)}, we see that this simple change in weighting can cause a significant
change in the expected cascade size $\rho$. In Fig.~\ref{fig:4}{\color{blue}(a)} the nonclustered graph produces
a larger value of $\rho$ than the clustered graph at every value of the threshold distribution mean $R$.
However, in Fig.~\ref{fig:4}{\color{blue}(b)} this trend is reversed in the region from approximately $R=0.26$
upward. Based on this observation, we submit that weighted models such as the one provided here may offer new
insights into the effects of clustering and decision bias in cascades on social networks \footnote{A model of
cascades on weighted multiplex networks has been proposed recently by Ya\u{g}an and Gligor \cite{yagan:12}.
The application of this model to clique-based graphs may also provide novel insights.}. We leave the analysis
and modification of this weighted model open to further investigation.

%*********************************
\section{Conclusion} \label{sec:5}
%*********************************

We have extended Gleeson and Cahalane's \cite{gleeson:07} analytical approach to modeling cascading phenomena
on configuration model graphs to the highly clustered clique-based graphs defined by Gleeson in
\cite{gleeson:09a}. An analytical expression for the expected cascade size and a first-order cascade
condition have been derived. The use of the generalized response function mechanism in these expressions permits
their application to a range of processes that includes site and bond percolation, $k$-core decomposition, and
Watts's threshold model.

We have validated our approach against numerical simulations of bond percolation and Watts's model. In addition,
we have proposed a modification of Watts's model that employs the unique structure of clique-based graphs in an
investigation of the role of group influence in processes of social contagion. This presents rich ground
for further investigation. The analytical framework provided by us here may be useful for such studies.

Perhaps the most significant aspect of our contribution is the derivation of a closed-form expression
for the steady-state fraction of active vertices inside a clique of arbitrary size. We anticipate that this
expression will find additional applications outside of the current setting.

Finally, there are a number of significant challenges that we have yet to address in our broad study of cascades
on clustered graphs. We have now provided approaches for a class of monotone binary-state processes on both
edge-triangle graphs \cite{hackett:11} and clique-based graphs; there are two directions in which we would like
to extend this work. First, we would like to modify our techniques to investigate nonmonotone processes. The
groundwork for this has been laid in \cite{gleeson:11}. Second, we would like to investigate cascades on a
more sophisticated class of highly clustered graphs than those dealt with so far. Such classes have been
described in \cite{karrer:10}.

%***********************
\begin{acknowledgements}
Discussions with Peter Sheridan Dodds are gratefully acknowledged. This work was funded by Science Foundation
Ireland under Programs No. 06/IN.1/I366, No. 11/PI/1026, and No. MACSI 06/MI/005.
\end{acknowledgements}
%***********************

%****************************
\appendix*
\section{Clusters in damaged graphs} \label{app:a}
%****************************

The results illustrated in Fig.~\ref{fig:3} demonstrate the equivalence of the approach
to bond percolation provided in \cite{gleeson:09a} and the corresponding approach provided
here. By working through the equations of \cite{gleeson:09a} and those of this paper, one
may show that the match between the two approaches hinges on the following equation:
\begin{equation}
\sum_{m=1}^{c}P(m|c)x^{m-1} = \sum_{m=0}^{c-1}R_m^{c-1}q^m,
\label{eq:A1}
\end{equation}
where $x=1-G_0^{c-2}$ and $q=1-\phi_b$. On the left-hand side of Eq.~(\ref{eq:A1})
$P(m|c)$ is the probability that in a $c$-clique that has been damaged by the removal
of its edges (each with independent probability $q$) a connected cluster of $m$
vertices (not necessarily an $m$-clique) remains.

In \cite{newman:03a} the probability $P(m|c)$ was evaluated iteratively using a
recursive formula; an explicit formula for $P(m|c)$ was not provided. By making
use of Eq.~(\ref{eq:A1}) we can now write an explicit formula for $P(m|c)$.
\\

Applying Eq.~(\ref{eq:20}) allows us to expand the right-hand side of
Eq.~(\ref{eq:A1}) and thereby rewrite it as
\begin{align}
& \sum_{m=1}^{c}P(m|c)x^{m-1} = \nn \\
& \quad \sum_{|l| \leq c-1}{c-1 \choose l_1}q^{l_1}\prod_{i=2}^{c-1} B_{l_i}^{n_{c-1,i}}(\theta_{c-1,i}) q^{l_i} \nn \\
& \quad \quad \times \sum_{j=0}^{l_1}{l_1 \choose j}(-1)^{j}x^{c-1-l_1+j}.
\label{eq:A2}
\end{align}

To equate coefficients of powers of of $x$ on the left-hand side and right-hand side
of Eq.~(\ref{eq:A2}) we simply set $j=m-c+l_1$. This gives us the following expression
for the probability $P(m|c)$:
\begin{align}
P(m|c) = &\sum_{|l| \leq c-1}{c-1 \choose l_1}q^{l_1}\prod_{i=2}^{c-1} B_{l_i}^{n_{c-1,i}}(\theta_{c-1,i}) q^{l_i} \nn \\
& \quad \quad \times{l_1 \choose m-c+l_1}(-1)^{m-c+l_1}.
\label{eq:A3}
\end{align}
One may easily verify that Eq.~(\ref{eq:A3}) satisfies the normalization condition $\sum_{m=1}^{c}P(m|c)=1$.

%************************************
\bibliography{clique_based_graphs_v4}
%************************************

\end{document}